\documentclass{ws-procs9x6}

\def\d0{D\O}
\def\D0{D\O}

\def\etmisv {\mbox{${\hbox{${\vec E}$\kern-0.6em\lower-.1ex\hbox{/}}}_T$}}
\def\etmis  {\mbox{${\hbox{$E$\kern-0.6em\lower-.1ex\hbox{/}}}_T$}}

\def\ifmath#1{\relax\ifmmode #1\else $#1$}%
\def\TeV{\ifmmode {\mathrm{ Te\kern -0.1em V}}\else
                   \textrm{Te\kern -0.1em V}\fi}%
\def\GeV{\ifmmode {\mathrm{ Ge\kern -0.1em V}}\else
                   \textrm{Ge\kern -0.1em V}\fi}%
\def\MeV{\ifmmode {\mathrm{ Me\kern -0.1em V}}\else
                   \textrm{Me\kern -0.1em V}\fi}%
\def\GeVcc{\ifmmode {\mathrm{ \GeV/c^2}}\else
                   \textrm{Ge\kern -0.1em V/c$^2$}\fi}%
\def\MeVcc{\ifmmode {\mathrm{ \MeV/c^2}}\else
                   \textrm{Me\kern -0.1em V/c$^2$}\fi}%

\def\ttbar              {\mbox{$t\overline{t}$}}

\def\Aslash{\mbox{${\hbox{$A$\kern-0.55em\hbox{/}}}$}}
\def\pslash{\mbox{${\hbox{$p$\kern-0.45em\hbox{/}}}$}}
\def\to{\rightarrow}

\def\gesim{\,{\raise-3pt\hbox{$\sim$}}\!\!\!\!\!{\raise2pt\hbox{$>$}}\,}
\def\lesim{\,{\raise-3pt\hbox{$\sim$}}\!\!\!\!\!{\raise2pt\hbox{$<$}}\,}
\def\boldoverdot{\,{\raise6pt\hbox{\bf.}\!\!\!\!\>}}

\def\diag{\hbox{\diag}}

\def\doubleundertext#1{
{\undertext{\vphantom{y}#1}}\par\nobreak\vskip-\the\baselineskip\vskip4pt%
\undertext{\hbox to 2in{}}}
\def\inbox#1{\vbox{\hrule\hbox{\vrule\kern5pt
     \vbox{\kern5pt#1\kern5pt}\kern5pt\vrule}\hrule}}
\def\sqr#1#2{{\vcenter{\hrule height.#2pt
      \hbox{\vrule width.#2pt height#1pt \kern#1pt
         \vrule width.#2pt}
      \hrule height.#2pt}}}

\def\today{\ifcase\month\or
  January\or February\or March\or April\or May\or June\or
  July\or August\or September\or October\or November\or December\fi
  \space\number\day, \number\year}
\def\pmb#1{\setbox0=\hbox{#1}%
  \kern-.025em\copy0\kern-\wd0
  \kern.05em\copy0\kern-\wd0
  \kern-.025em\raise.0433em\box0 }
\def\sumprime_#1{\setbox0=\hbox{$\scriptstyle{#1}$}
  \setbox2=\hbox{$\displaystyle{\sum}$}
  \setbox4=\hbox{${}'\mathsurround=0pt$}
  \dimen0=.5\wd0 \advance\dimen0 by-.5\wd2
  \ifdim\dimen0>0pt
  \ifdim\dimen0>\wd4 \kern\wd4 \else\kern\dimen0\fi\fi
\mathop{{\sum}'}_{\kern-\wd4 #1}}
\begin{document}

\title{Top Quark Physics Results from \d0}

\author{John Ellison}

\address{University of California, Riverside\\
E-mail: john.ellison@ucr.edu}

\maketitle

\vspace*{-6mm}
\centerline{For the \d0\ Collaboration}
\vspace*{2mm}

\abstracts{
We describe recent measurements of the $\ttbar$ production cross section
and the mass of the top quark from the \d0\ Collaboration.}

\section{Introduction}

In this paper we describe some recent results on the production and
properties of top quarks obtained by the \d0\ Collaboration at the
Fermilab Tevatron Collider. New measurements of the top quark production cross
section have been made using 45--107~pb$^{-1}$ of data collected at
$\sqrt{s} = 1.96$~TeV. We also describe new measurements of the top
quark mass 
%and the polarization of $W$ bosons in top quark decay 
using 125~pb$^{-1}$ of data collected at $\sqrt{s} = 1.8$~TeV.

\section{\boldmath Measurements of the $\ttbar$ Production Cross Section}

Production of $\ttbar$ pairs at the Tevatron at $\sqrt{s} = 1.96$~TeV
proceeds via quark-antiquark annihilation ($\approx 85$\%) and
gluon-gluon fusion ($\approx 15$\%) and the theoretical prediction for
the production cross section is
6.8~pb$^{-1}$~[\refcite{kidonakis},\refcite{cacciari}]. Comparisons of
the prediction with experimental measurements are an important test of
our understanding of the QCD production mechanisms. Because of the
unusually high mass of the top quark, it has been suggested that the
top quark may play a special role in symmetry breaking and
that non-Standard-Model physics may show up as anomalies in top quark
production and decay. Precise measurements of the $\ttbar$ cross
section utilizing different top decay channels are therefore an
important goal.

Recent measurements made by the \D0\ Collaboration include the
dilepton decay channels $ee$, $\mu \mu$, and $e \mu$ and the lepton
plus jets channels $e$+jets and $\mu$+jets. The dilepton channels have
small branching fractions but the signal is relatively free
of background. Events are required to contain two charged leptons,
significant missing transverse energy ($\etmis$) consistent with the
decay $t \to Wb \to \ell
\nu b$, and at least two jets. The results are summarized in
Table~\ref{table1}.
\begin{table}[ph]
\tbl{Event yields, estimated backgrounds, and expected numbers of 
signal events for the dilepton analyses.}
{\footnotesize
\begin{tabular}{@{}lccc@{}}
\hline
{} &{} &{} &{} \\[-1.5ex]
{} & $ee$ & $\mu\mu$ & $e\mu$ \\[1ex]
\hline
{} &{} &{} &{} \\[-1.5ex]
Int. Luminosity (pb$^{-1}$) & 107 & 90.4 & 97.7\\[1ex]
Background & $0.6 \pm 0.5$ & $0.7 \pm 0.4$ & $0.6 \pm 0.4$ \\[1ex]
Expected signal & $0.6 \pm 0.1$ & $0.5 \pm 0.1$ & $1.7 \pm 0.3$ \\[1ex]
Signal + bkgd & $1.2 \pm 0.5$ & $1.2 \pm 0.5$ & $2.3 \pm 0.5$ \\[1ex]
Data & 2 & 0 & 3 \\[1ex]
\hline
\end{tabular}\label{table1} }
\vspace*{-13pt}
\end{table}

The lepton + jets channels have larger branching fractions but also
much large backgrounds compared with the dilepton channels. Events are
required to have a high-$p_T$ electron or muon, significant missing
$E_T$, and at least three or four jets, consistent with $\ttbar \to Wb~Wb \to
\ell \nu j j b b$. To further separate the signal from the $W$ + jets background, the events are required to pass one of the following tagging algorithms:\\

\noindent {\it Topological Tag} 
($\ell \nu + \ge 4$~jets, 92~pb$^{-1}$):
Aplanarity $A > 0.065$, $H_T^{all} = p_T(W) + \sum_{jets} E_T 
> 180$~GeV ($e$+jets)~or~$> 220$~GeV ($\mu$ + jets)\\

\vspace*{-2mm}
\noindent {\it Soft Muon Tag}
($\ell \nu + \ge 4$~jets, 92~pb$^{-1}$):
$A > 0.004$, $H_T^{all} > 110$~GeV, and a
muon with $p_T > 4$~GeV, $| \eta | < 2.0$ within 
$\Delta R < 0.5$ of a jet\\

\vspace*{-2mm}
\noindent {\it Secondary Vertex Tag (SVT)}
($\ell \nu + \ge 3$~jets, 45~pb$^{-1}$):
Secondary vertex with transverse decay length satisfying 
$L_{xy} / \sigma(L_{xy}) > 5$\\

\vspace*{-2mm}
\noindent {\it Charged Signed Impact Parameter Tag (CSIP)}
($\ell \nu + \ge 3$~jets, 45~pb$^{-1}$):
Two tracks with charge signed impact parameter 
significance $>$ 3; or three tracks with CSIP significance $>$ 2\\

\begin{figure}[ht]
\begin{tabular}{l l}
       \epsfxsize = 5.25cm
       {\epsffile{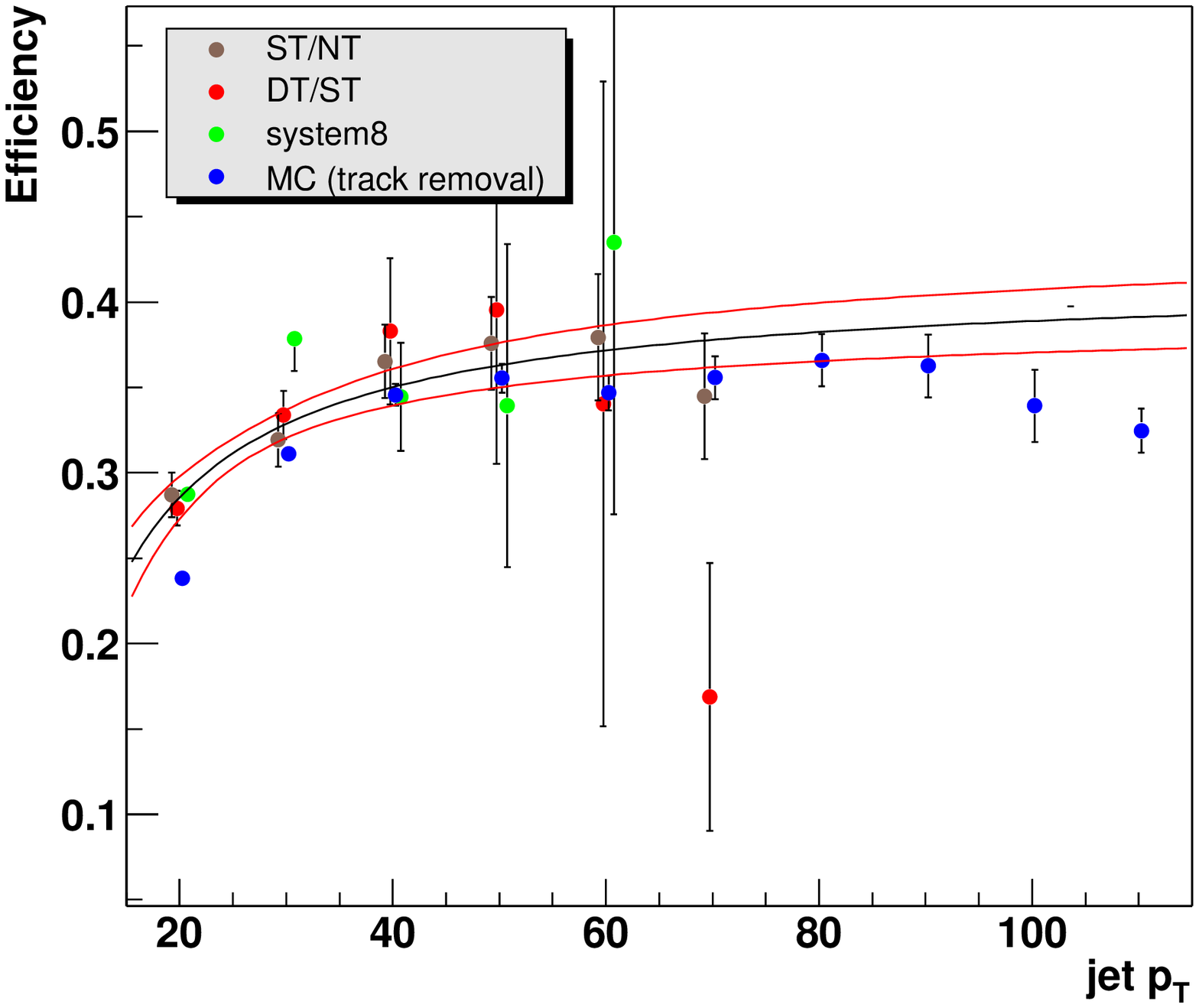}} &
       \epsfxsize = 5.25cm
       {\epsffile{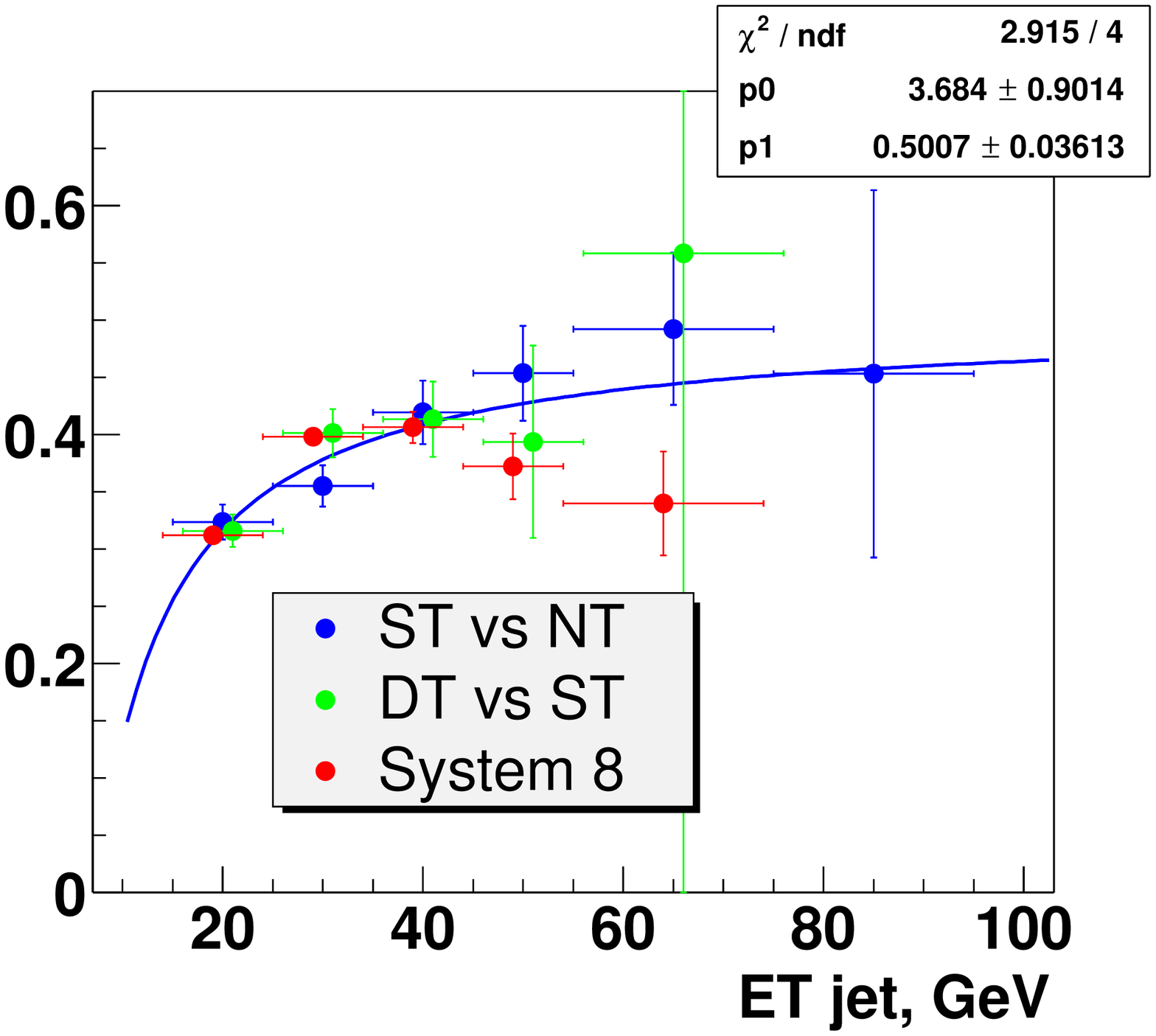}}
\end{tabular}
\caption{Efficiency for b-tagging as a function of jet $p_T$ for the 
CSIP (left) and SVT (right) tagging algorithms. 
\label{fig:eff_tag}}
\end{figure}

\begin{figure}[ht]
\epsfysize=4.75cm
\centerline{\epsfbox{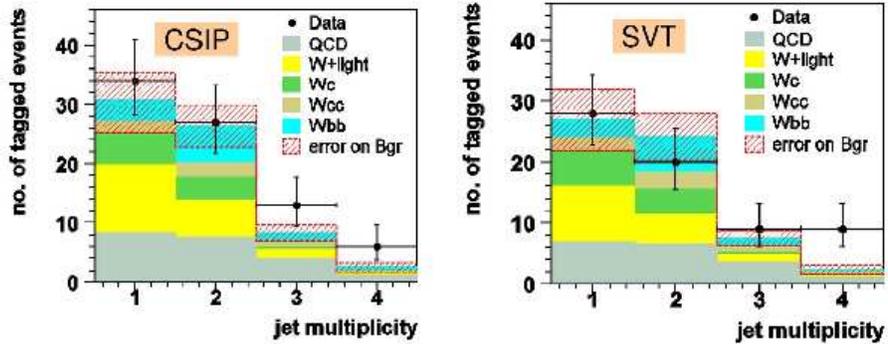}}   
\caption{Distribution of the observed number of jets per event for the 
CSIP and SVT tagged samples. 
\label{fig:njet}}
\end{figure}

Here we briefly discuss results from the SVT and CSIP tag
analyses. The $b$-jet tagging efficiency as a function of jet $p_T$ is
shown in Fig.~\ref{fig:eff_tag}, where the measured efficiency
(points) is compared to the expectation from Monte Carlo (line). The
resulting distribution of number of jets per event for the two
analyses after all event selection requirements is shown in
Fig.~\ref{fig:njet}. A clear excess of events in the signal region
($\ge 3$~jets) can be seen.

\begin{figure}[ht]
\epsfysize=8cm
\centerline{\epsfbox{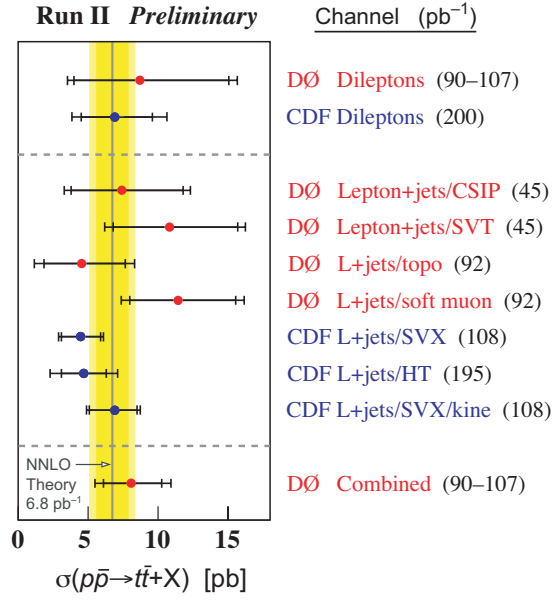}}   
\caption{Measurements of the $\ttbar$ production cross section at 
$\sqrt{s} = 1.96$~TeV. 
\label{fig:XS}}
\end{figure}

The cross section measurements obtained from the above analyses are
shown in Fig~\ref{fig:XS}. As can be seen, the measurements are
consistent with the theoretical predictions. The combined \d0\ cross
section is $\sigma(\ttbar) = 
8.1~^{+2.2}_{-2.0} \; (\mathrm{stat}) 
~ ^{+1.6}_{-1.4} \; (\mathrm{syst}) 
~ \pm 0.8 \; (\mathrm{lum})$~pb.

\section{Improved Measurement of the Top Quark Mass}

A precise measurement of the top quark mass ($M_t$) is important because of
the dependence of the electroweak precision observables on $M_t$. The
prediction of the Higgs boson mass from fits to the electroweak data is
particularly sensitive to $M_t$.

A new method based on event probability has been used to obtain an
improved measurement of the top quark mass by \d0. The new method uses
the Run I lepton + jets data selected by requiring an isolated
electron or muon with $E_T > 25$~GeV and pseudorapidity $\left|
\eta^e \right| < 2.0$, $\left| \eta^\mu \right| < 1.7$, at least four
jets with $E_T > 25$~GeV and $\left| \eta^{jet} \right| < 2.0$,
$\etmis > 20$~GeV, and requiring  the $W$ boson decay
products to satisfy $E_T^{lep} + \etmis > 60$~GeV, and
$\left| \eta_W \right| < 2.0$.

\begin{figure}[ht]
\begin{tabular}{c c}
       \epsfxsize = 5.4cm
       {\epsfbox{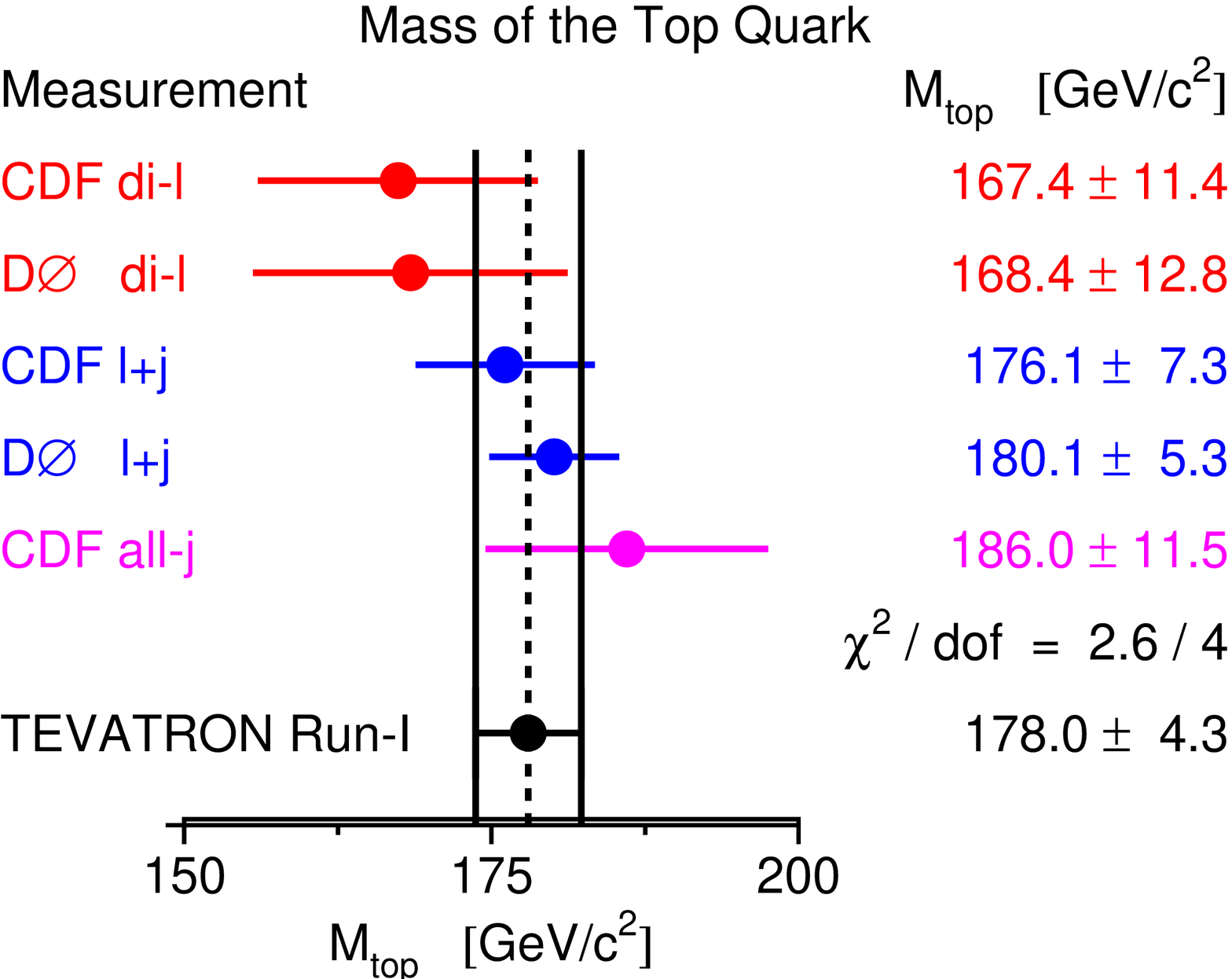}} &
       \epsfxsize = 5.4cm
       {\epsfbox{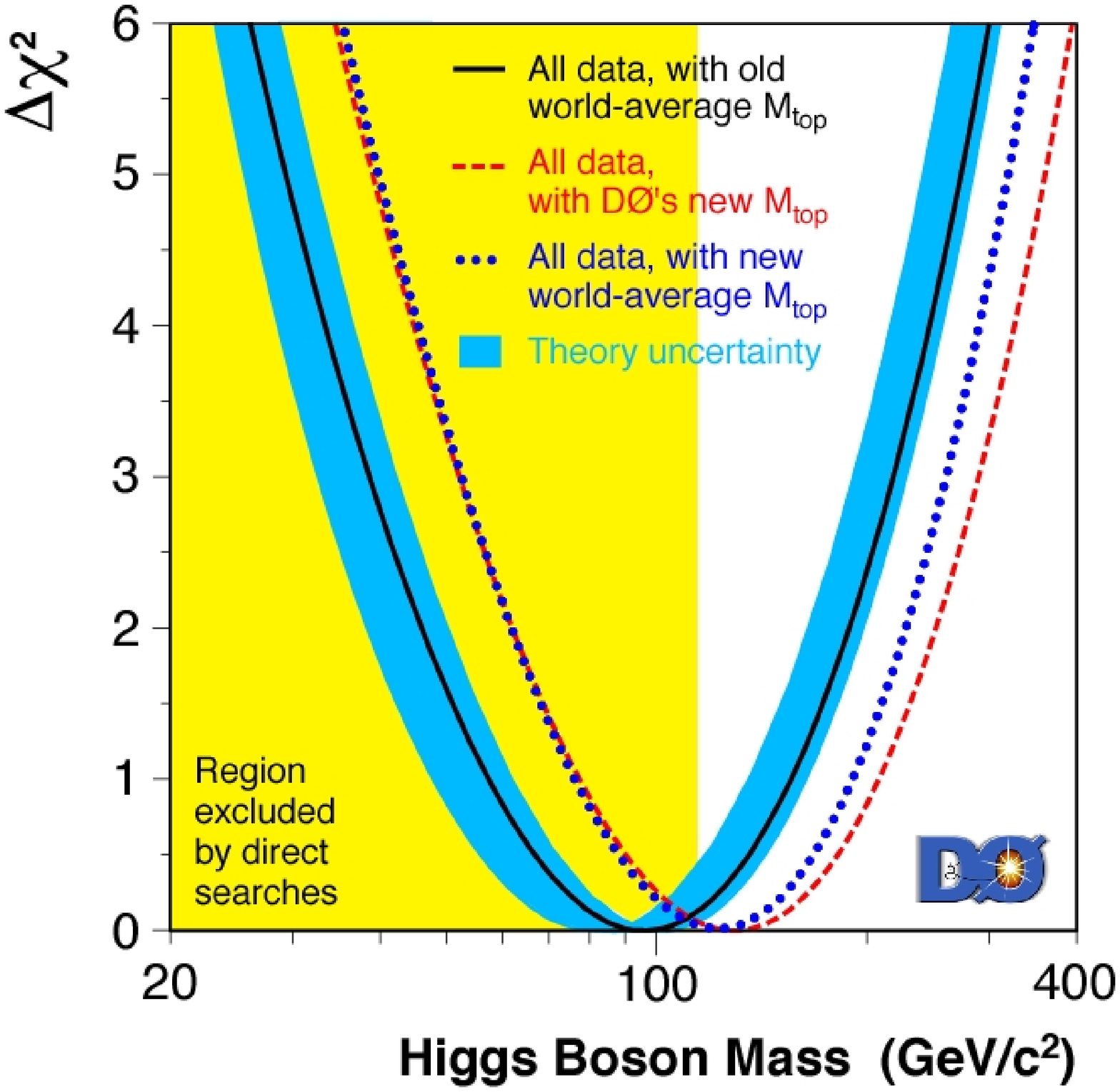}} 
\end{tabular}
\caption{Measurements of the top quark mass (left) and $\Delta \chi^2$ vs. $m_H$ from fits to the EW data (right). 
\label{fig:mt_mh}}
\end{figure}

A probability density is defined for each event, which uses all
measured quantities from each reconstructed event:
\begin{equation*}
P(x, M_t) = \frac{1}{\sigma(M_t)} 
\int d\sigma(y, M_t) dq_1 dq_2 f(q_1) f(q_2) W(y,x) 
\end{equation*}

\noindent
where $d\sigma(y, M_t)$ is the differential cross section calculated
from the leading-order (LO) matrix element and phase space factor, $f(q_1),
f(q_2)$ are the parton distribution functions, and $W(x,y)$ is the
transfer function which relates partonic four-momenta ($y = (p_1, p_2,
... p_n)$) to the measured four-momenta $x$ of the final state
particles and jets. Because of the use of the LO matrix element, a
further requirement that the selected events must contain exactly four
jets is imposed. The ambiguity of the jet-parton correspondence is
handled by summing over all twelve assignments of the four jets to the
partons ($b \bar b q_3 q_4$). This means that all combinations are
used correctly, whereas in the previous mass measurement
method~[\refcite{mt_run1}] only the solution with the best $\chi^2$
was chosen. The total event probability is defined as
\begin{equation*}
P(x; c_1, c_2,M_t) = c_1 P_{t \bar t}(x; M_t) + c_2 P_{bkgd}(x)
\end{equation*}

\noindent
where $c_1, c_2$ are constants and $P_{bkgd}$ is the probability
density for background, defined using the LO matrix element for the
$W$ + 4~jets process. A likelihood function is formed using $P(x; c_1,
c_2,M_t)$ which is then maximized to obtain $M_t, c_1$ and $c_2$.  An
additional cut on the background probability $P_{bkgd} < 10^{-11}$ is
applied to increase the signal purity, resulting in 22 events
surviving the final cuts. The result for the fitted top quark mass
is $M_t = 180.1 \pm 3.6 \; (\mathrm{stat}) ~ \pm 3.9 \;
(\mathrm{syst}) = 180.1 \pm 5.3$~GeV$/c^2$. The improvement in
statistical uncertainty compared with the previous result is
equivalent to a factor of 2.4 more data. Using this new result in
place of the previous \d0\ lepton + jets result gives a world average
top quark mass of $178.0 \pm 4.3$~GeV$/c^2$ and changes the best-fit
Higgs mass from 96~GeV$/c^2$ to approximately 125~GeV$/c^2$ (see
Fig.~\ref{fig:mt_mh}).

\end{document}